# Single Photon Source from a Nanoantenna-Trapped Single Quantum Dot

Quanbo Jiang,[1] Prithu Roy,[1] Jean-Benoît Claude,[1] Jérôme Wenger[1,*]

[1] *Aix Marseille Univ, CNRS, Centrale Marseille, Institut Fresnel, AMUTech, Marseille, France*

[*] *Corresponding author:* jerome.wenger@fresnel.fr

**Abstract**

Single photon sources with high brightness and subnanosecond lifetimes are key components for quantum technologies. Optical nanoantennas can enhance the emission properties of single quantum emitters, but this approach requires accurate nanoscale positioning of the source at the plasmonic hotspot. Here, we use plasmonic nanoantennas to simultaneously trap single colloidal quantum dots and enhance their photoluminescence. The nano-optical trapping automatically locates the quantum emitter at the nanoantenna hotspot without further processing. Our dedicated nanoantenna design achieves a high trap stiffness of 0.6 fN/nm/mW for quantum dot trapping, together with a relatively low trapping power of 2 mW/µm². The emission from the nanoantenna-trapped single quantum dot shows 7× increased brightness, 50× reduced blinking, 2× shortened lifetime and a clear antibunching below 0.5 demonstrating true single photon emission. Combining nano-optical tweezers with plasmonic enhancement is a promising route for quantum technologies and spectroscopy of single nano-objects.

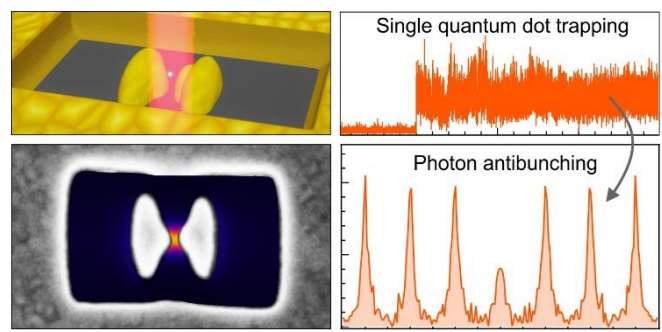

Figure for Table of Contents

The development of bright, fast, and controllable single photon sources is a major challenge for quantum information and communication processing.[1–5] In an homogeneous environment, single photon quantum emitters suffer from low brightness, long fluorescence lifetimes and limited collection efficiencies. The introduction of optical nanoantennas holds great promises to improve the emission properties and circumvent all the shortcomings set by the diffraction limit.[6–8] Impressive results have been achieved over the recent years to reach several orders of magnitude enhancement in the emission brightness,[9–11] photokinetic rates,[12–14] and collection efficiencies.[15–19]

A major requirement for optical nanoantennas is that the quantum emitter must be positioned at the antenna hotspot with nanoscale precision in order to achieve the highest enhancement. Many approaches have been explored, relying on random dispersion,[20–22] two-step electron-beam lithography,[15,23] two-step nanofabrication,[17,24] DNA origami,[9,25] DNA self-assembly,[14,26] or atomic force microscopy.[27] However, none of these techniques offers the simplicity and flexibility of plasmonic nano-optical tweezers where the trapped object is automatically positioned at the antenna hot spot.[28–32] Using the intense field gradients in plasmonic nanoantennas,[33–38] it becomes possible to manipulate single quantum emitters such as NV centers,[39,40] quantum dots,[41–43] erbium-doped nanocrystals,[44,45] or even a single atom.[46] For a source to be considered into the quantum regime with single photon emission, it must verify the antibunching condition stating that its second-order intensity correlation at zero lag time $g^{(2)}(0)$ must not exceed the 0.5 threshold.[47,48] Despite intense research, photon antibunching with $g^{(2)}(0) < 0.5$, which gives the clear quantum signature of single photon emission, has remained elusive for a trapped quantum nano-object.

Here, we introduce a dedicated plasmonic nanoantenna design to trap single colloidal quantum dots (QDs) acting as single photon sources. Our nanogap antenna provides simultaneously strong field gradients in the infrared for plasmonic trapping together with QD photoluminescence enhancement. We demonstrate nano-optical trapping of single CdSe/ZnS quantum dots of 11 nm diameter and achieve a trap stiffness of 0.6 fN/nm/mW which is the highest reported to date for QD trapping.[41,42] The infrared trapping intensity can be as low as 2 mW/µm², which is 8× lower than the previous state-of-the-art,[41,42] allowing to significantly mitigate the unwanted Joule heating and thermal-related effects.[49,50] The electromagnetic near-field coupling between the single quantum dot and the nanoantenna is thoroughly investigated by measuring the photoluminescence (PL) dynamics on a broad timescale from picoseconds to seconds. Most importantly for quantum technology applications, we show a clear antibunching of $g^{(2)}(0) = 0.41 \pm 0.03$, clearly below the 0.5 threshold and currently limited by the emission properties of the quantum dot source. By providing a versatile solution to the localization problem of optical nanoantennas, the plasmonic trapping of single quantum objects shows high potential to develop bright ultrafast single photon sources,[2,7,13,39] and improve the spectroscopy analysis of single nano-objects.[51,52]

Figure 1a shows a scheme of our microscope combining a 1064 nm infrared laser for trapping with a 635 nm laser for PL excitation. Two avalanche photodiodes in a Hanbury-Brown-Twiss configuration enable a detailed analysis of the PL dynamics. The plasmonic nanogap antennas are milled by focused ion beam in a 100 nm thick gold film (Fig. 1b,c and S1,S2) with a 22 or 34 nm gap size to accommodate for the quantum dot diameter. We have selected this antenna design so as to yield strong intensity gradient at 1064 nm while keeping a high intensity enhancement at 635 nm (Fig. 1b). As compared to the double nanohole design,[53,54] the intensity enhancement for the 635 nm laser in the center of the gap is about three times higher which directly improves the photoemission of the trapped quantum dot (Fig. S3). While doing preliminary trapping experiments on nanoantennas using two gold hemispheres like the ones used in [10,55], we observed trapping events of multiple nanoparticles, which were attributed to the intensity hot spots occurring on the opposite extreme edges of the dimer antenna. The elliptical shape of the gold antenna with a protrusion to define the gap region solves this issue by promoting the hotspot in the nanogap center and quenching the intensity gradients on the extreme edges of the antenna. The design optimization is a trade-off between maximizing the responses at 635 and 1064 nm (Fig. S4 and S5). Because of the large spectral shift between these wavelengths, there is no specific set of parameters that is optimal for both wavelengths. We choose here to maximize the PL excitation gain at 635 nm while keeping fixed the long 220 nm edge of the antenna in order to have a single hotspot for trapping. The presence of the QD into the gap may affect the local intensity enhancement by introducing additional damping mechanisms for the plasmonic resonance. To account for this effect, we have performed numerical simulations where the QD is modeled by a 10 nm sphere of 2.47 refractive index positioned in the center of the gap. Figure S5 indicates that the presence of the QD lowers the net intensity enhancement experienced by the QD by 2 to 3-fold as compared to simulations performed without the QD, so that the expected intensity gain at 635 nm is 22×.

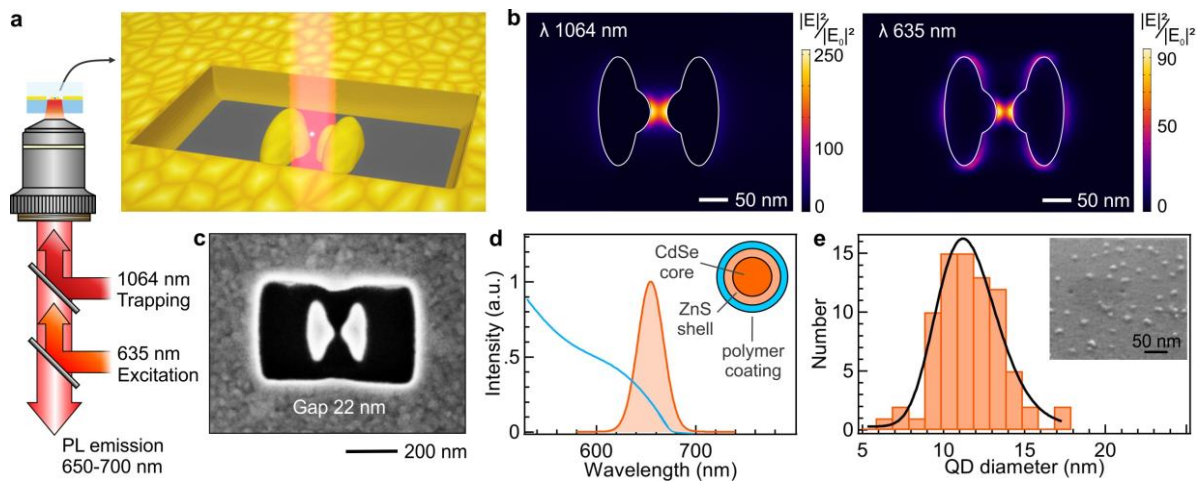

**Figure 1.** Plasmonic nano-optical trapping of a single quantum dot. (a) Scheme of the experimental setup with a 1064 nm laser for trapping and a 635 nm laser for photoluminescence excitation. (b) Numerical simulations of the intensity enhancement at 1064 nm and 635 nm in the gap nanoantenna, 5 nm above the glass-gold interface. The laser polarization is parallel to the gap. (c) Scanning electron microscopy image of a nanoantenna with 22 nm gap size. (d) Absorption (blue) and emission (orange) spectra of a core/shell QD. (e) QD size distribution measured by scanning electron microscopy, the inset shows a typical SEM image of QDs dispersed on an ITO-coated coverslip viewed at 52° incidence.

We use commercial quantum dots (Thermo Fisher Scientific Qdot™ 655 ITK™ Q21321MP) with a CdSe core and a ZnS shell. The quantum dots (QDs) are coated with a polymer layer with a carboxyl surface functionalization making them water soluble. The peak emission occurs at 655 nm (Fig. 1d) and the quantum efficiency is estimated at 60% (see Supporting Information section S1). We have checked that the infrared laser does not affect the PL brightness and lifetime (Fig. S6) and that the two-photon absorption of the QD at 1064 nm remains negligible in our conditions even with the presence of the nanoantenna (Fig. S7). The QD size distribution is determined experimentally from scanning electron microscopy images with an average diameter of 11 nm and a standard deviation of 2 nm (Fig. 1e).

Trapping of single QDs is demonstrated in Fig. 2a,b using nanoantennas of 22 or 34 nm gap size. In these experiments, the 1064 nm infrared laser is switched on after 3 s at 2 or 3 mW/µm² while the 635 nm laser is constantly present at 0.5 µW. Monitoring the PL signal allows to work on a nearly-zero background and unequivocally distinguish when a QD is trapped and when it escapes from the trap. The background noise level is determined at 0.8 kcnts/s from the intensity detected while no QD is present in the trap (Fig. 2a). The dark count of our avalanche photodiodes is 0.1 kcnts/s and no backscattered signal is detected from the 1064 nm beam itself so our main source of background noise is a residual 0.7 kcnts/s coming from the incomplete filtering of the 635 nm laser back-reflection and potentially some PL from the antenna itself or other optical elements on our setup. When a QD is trapped in the nanoantenna, a clear jump of the photoluminescence signal above the background is observed. Besides, the PL signal returns to the background level when the QD escapes from the trap. This indicates the single QD nature of the trapping events and shows that the QD is not electrostatically adsorbed on the metal surface. Longer trapping events are observed with the smaller 22 nm gap size as compared to the 34 nm gap, with a duration increasing with the trapping laser power. Stable trapping of a single QD for several seconds is achieved with a relatively low continuous wave infrared intensity of 2 mW/µm², which is 8× lower than previous reports of QD nano-optical trapping.[41,42] Importantly, using a lower infrared power allows to mitigate the temperature elevation around the nanostructure and avoid unwanted thermal-related effects. Using the technique developed in [49,50], we measure a temperature increase of +7.5°C at 2 mW/µm² for our 22 nm gap antenna (Fig. S8).

To further confirm plasmonic trapping and enhancement from the nanogap region, we rotate the polarizations of the 1064 and 635 nm lasers (Fig. 2c). No trapping is observed when the 1064 nm polarization is perpendicular to the gap, which shows that plasmonic excitation of the nanogap hotspot is mandatory for our trapping experiments. Moreover, the PL signal is decreased by about 2× when the 635 nm laser polarization is turned perpendicular to the gap. While numerical simulations predict stronger attenuations (Fig. S5), the reason why the PL signal is only reduced by 2× remains unclear. This may be related to the degenerate nature of the QD dipole transition moment,[23,56] or the QD blinking reduction in presence of the nanoantenna.

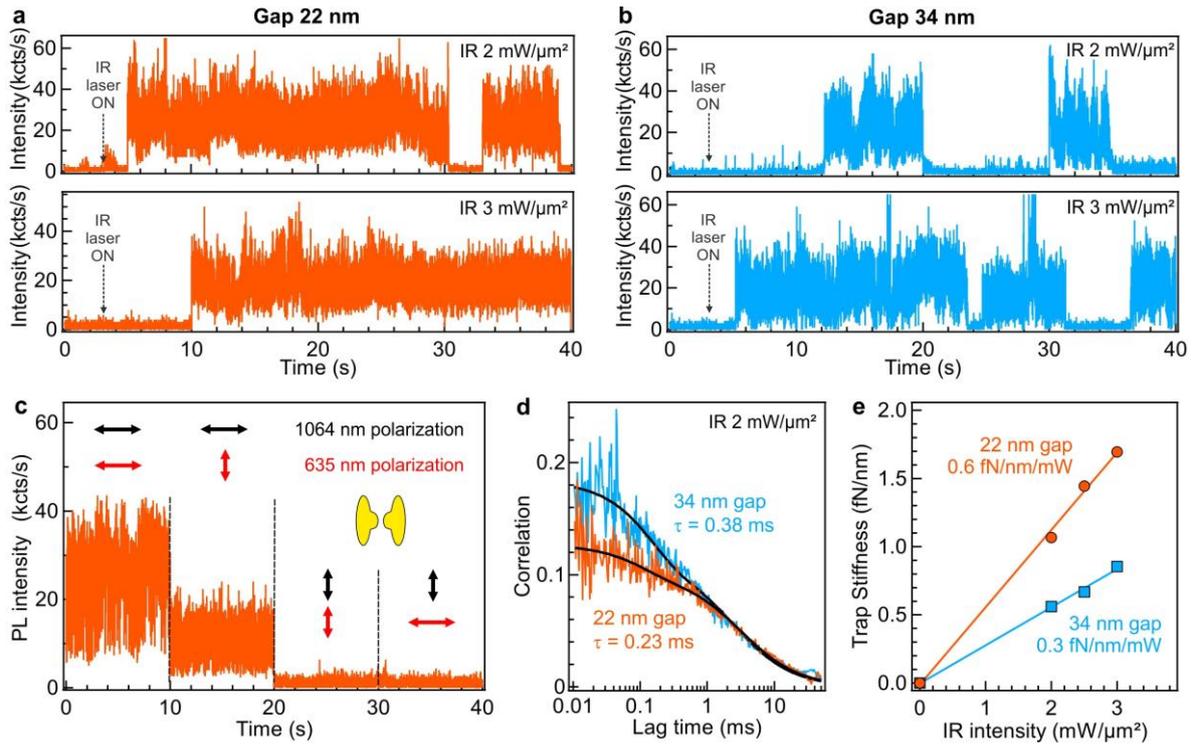

**Figure 2.** Nanoantenna trap characterization with single quantum dots. (a)-(b) PL intensity time traces of a trapped QD for different nanoantennas gap sizes and trapping laser powers. The sudden jumps on a nearly-zero background indicate trapping of a single QD. The binning time is 1 ms. (c) Polarization influence of the 1064 nm trapping laser (black arrows) and the 635 nm excitation laser (red arrows). (d) Correlation functions (orange and blue lines) and numerical fits (black lines) computed from the PL intensity recorded during trapping events for two different antenna gap sizes. The characteristic correlation times are indicated on the graph. (e) Evolution of the trap stiffness deduced from correlation data as a function of the infrared trapping power. The trap stiffness normalized by the infrared power is determined from the linear fits.

To quantify the trap performance of our nanoantennas with QDs, we use the correlation approach described in our earlier works.[53,57] Briefly, the temporal correlation of the PL intensity recorded during a trapping event is computed (Fig. 2d). This correlation decays as $\exp(-2t/\tau)$ where the characteristic correlation time $\tau = \gamma/\kappa$ corresponds to the ratio of the Stokes drag coefficient $\gamma$ by the trap stiffness $\kappa$. We use Faxen's law [58,59] to compute $\gamma$ and then determine $\kappa$ from the measured correlation time (see Supporting Information section S1 for details). Figure 2e summarizes our results for the 22 and 34 nm gap. The trap stiffness increases linearly with the infrared laser power, with a slope of 0.6 fN/nm/mW for the 22 nm gap antenna for trapping 11 nm QDs. This value is about 50% higher than the 0.42 fN/nm/mW reported while trapping individual silica-coated QDs using a bowtie nanoaperture.[41] It is important to take into account the differences in polarizabilities between the trapped objects while comparing plasmonic trapping experiments. Supplementary data using polystyrene nanoparticles (Fig. S9) experimentally assess the improved trapping performance of the gap nanoantennas as compared to the double nanohole structure. For a 20 nm diameter polystyrene nanoparticle, we estimate a normalized trap stiffness of 1.1 fN/nm/mW, which favorably compares to the reported normalized stiffness of 0.1 fN/nm/mW for a double nanohole,[60] 0.12 fN/nm/mW for a resonant bowtie aperture,[36] 0.25 fN/nm/mW for connected nanohole arrays,[61] 0.36 fN/nm/mW for a coaxial aperture,[62] and 8.65 fN/nm/mW for asymmetric split-rings metamaterial.[63] It is noteworthy to mention that our stiffness values benefit from the recent characterization of the role of the surfactant and the thermophoretic force in plasmonic trapping.[57] 2 mM sodium dodecyl sulfate (SDS) surfactant ensures that the carboxylate-modified QD display a thermophilic behavior (Fig. S10). These conditions correspond to the situation where the thermophoretic effect strengthens the trap potential and further eases the trapping of nanoscale objects.[57]

Having demonstrated the trapping of a single QD in the nanogap antenna, we now investigate the QD PL dynamics and its modification in presence of the nanoantenna. All the data in Fig. 3 are taken with a single QD either trapped in the 22 nm gap antenna (orange traces) or fixed on a flat glass coverslip (blue traces) to serve as a reference. Both cases use the same 0.5 µW average power and 5 MHz repetition rate at 635 nm to excite the QD. The antenna presence significantly affects and improves several aspects of the PL dynamics. The average PL intensity in the antenna trap is 20 kcts/s (Fig. 3a,c) and is enhanced by 7× as compared to the 3 kcts/s reference average intensity on glass (Fig. 3b). Considering the ~60% quantum yield of the QDs and the 22 nm gap size, this PL enhancement value is quite comparable with earlier works on fluorescence enhancement using organic fluorescent dyes with similar quantum yield and 23 nm gap.[25] The 7× PL enhancement is about one third of the 22× excitation intensity gain predicted by numerical simulations for the linear polarization (Fig. S5). This reduction is expected due to the 3D rotation of the QD inside the trap.[14] Simultaneously to the PL enhancement, the QD blinking is strongly suppressed by the presence of the nanoantennas (Fig. 3a,b, S11). The correlation amplitude in the hundreds of ms timescale is reduced more than 50× as compared to the glass reference (Fig. 3d), we use this value as figure of merit for the blinking reduction. Besides, the on- and off- state probabilities are modified (Supporting Information Fig. S11), leading to a more stable PL emission in presence of the gold nanoantennas. This observation of QD blinking reduction goes in line with previous reports using conductive materials.[22,23,64–67] Both the higher brightness of the on-state and the strongly suppressed blinking contribute to improve the PL emission of the QD trapped in the antenna.

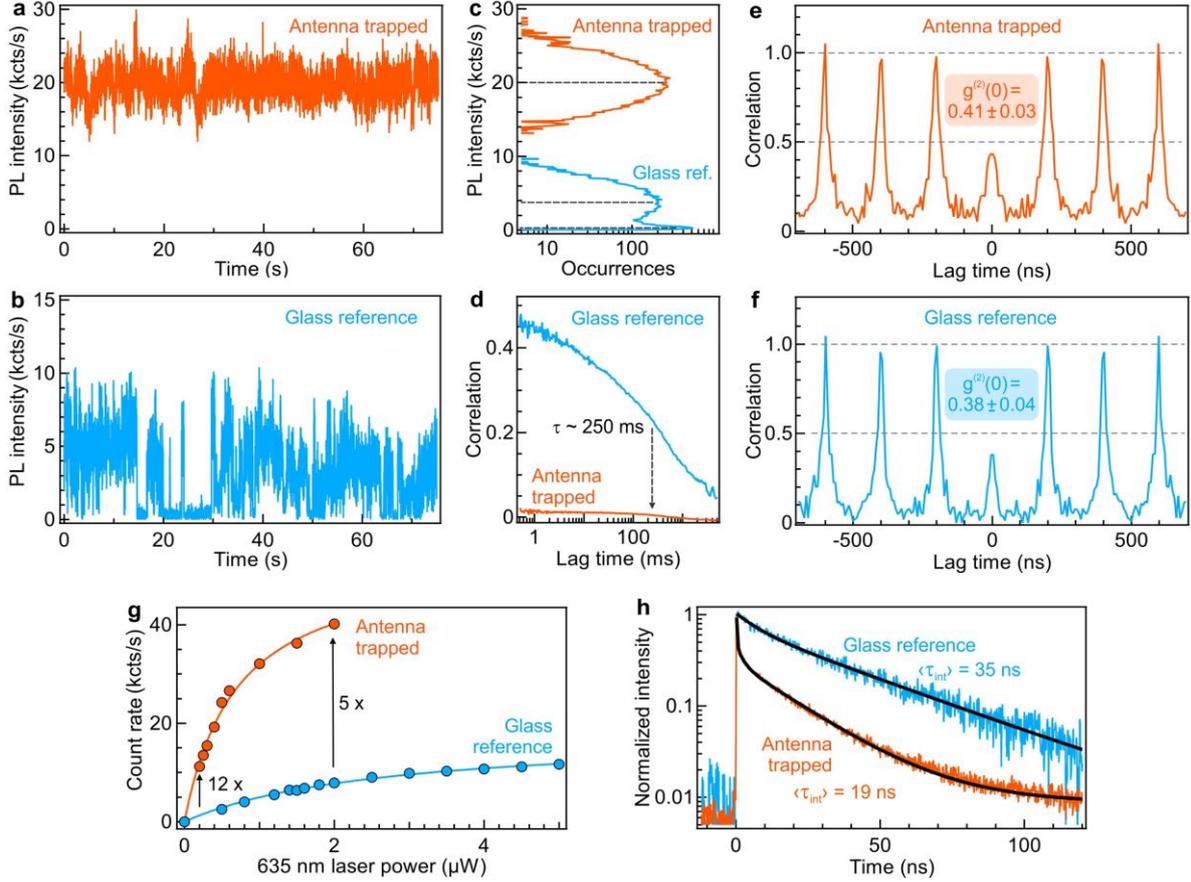

**Figure 3.** Photophysics of single QD trapped in a 22 nm nanogap antenna. (a,b) PL intensity time traces with 10 ms bin time of a single QD trapped in the gap antenna (a) or fixed on a glass coverslip (b). The 635 nm excitation power is constant at 0.5 µW average power with 5 MHz repetition rate and the infrared trapping intensity is 3 mW/µm². (c) Photon count histogram for the time traces in (a,b). (d) Correlation function in the ms to s timescale for the time traces in (a,b). (e,f) Hanbury-Brown-Twiss correlation of a single QD, the antibunching with $g^{(2)}(0) < 0.5$ at zero lag time is the signature of single-photon emission. This data is averaged out of 7 individual trapping experiments of at least 30 s each. (g) Dependence of the PL intensity with the 635 nm excitation power measured at the entrance port of the microscope. (h) PL lifetime decays and intensity-averaged PL lifetimes determined from the numerical fits. The fit results are detailed in the Supporting Information Tab. S1.

Photon antibunching is the main signature of experiments involving a single quantum emitter.[47] From the PL time traces in Fig. 3a,b, we compute the second order Hanbury-Brown-Twiss intensity correlation $g^{(2)}(\tau)$ (Fig. 3e,f).[48] The coincidence between our two avalanche photodiodes is maximal with $g^{(2)} = 1$ at lag times multiple of 200 ns, corresponding to the 5 MHz repetition rate of our 635 nm laser. At zero lag time, we observe a clear antibunching with $g^{(2)}(0) \cong 0.4$ for both the QD trapped in the antenna and fixed on a glass coverslip. This correlation at zero lag time is well below the 0.5 threshold that determines the boundary between single and multiple quantum emitters, demonstrating a plasmonic antenna-enhanced triggered single-photon emission.[2,51,68] For some selected QDs, we find an even lower antibunching with $g^{(2)}(0) \approx 0.2$ (Fig. S12), yet this corresponds to less than 10% of all QDs tested. Importantly, it shows that the residual correlation found at zero lag time is limited by the QD sample and its biexciton quantum yield,[22,51] and is not related to the nanoantenna trap.

We report the influence of the 635 nm excitation power in Fig. 3g. Without the antenna, the PL saturates at intensities below 10 kcts/s, which limits the maximum single photon emission rate. Thanks to the plasmonic enhancement in the nanoantennas, the brightness can significantly overcome the 10 kcts/s level, providing more than 5x gain even in the saturation regime. However, the biexciton contribution may be an issue at high excitation powers above the PL saturation,[52] reducing the antibunching at zero lag times.[51] Therefore the enhancement at lower excitation powers (in the linear regime) is of major interest for single photon sources. Thanks to the nanoantenna, we achieve brightness above 20 kcts/s even at moderate excitation powers below 0.8 µW. Further reduction of the nanoantenna gap size and optimization of the design parameters should promote higher enhancement values.

Lastly, we monitor a shortening of the PL lifetime in the nanoantenna trap as compared to the glass reference (Fig. 3h). To rule out any additional effect that could be involved into this lifetime reduction, we check that the 1064 nm laser beam alone does not alter the PL lifetime (Fig. S6f) and that the temperature increase around the plasmonic trap does not modify the PL decay kinetics (Fig. S10c). The lifetime reduction once the QD is trapped in the nanoantenna is directly a consequence of the higher local density of optical states (LDOS) in the nanogap region, which includes contributions from both the enhanced radiative emission (Purcell effect) and the non-radiative energy transfer to the metal (quenching effect). Figure S13 shows additional PL decay traces reporting the influence of the antenna gap size and the trapping power. Shorter lifetimes are observed with narrower gaps, in agreement with the LDOS enhancement expectations. Moreover, the lifetime is further reduced when the QD is more localized in the nanogap region thanks to a higher trapping power and a larger trap stiffness. While significantly larger lifetime reductions have been reported for single QDs in contact with silver nanocubes,[13] we believe that the 22 nm gap in our case limits the lifetime reduction.

Altogether, we demonstrate plasmonic-enhanced single photon emission from a single QD trapped in a nanogap antenna. The nano-optical trapping ensures that the quantum emitter is automatically positioned at the nanoantenna hotspot, maximizing the PL enhancement. Thanks to the nanoantenna, we achieve >7× increased brightness, >50× reduced blinking amplitude and 2× lifetime shortening. All these features improve the properties of the single photon source. Our approach remains fully general, and the excellent trapping performance of our antenna design makes it suitable to manipulate other nanoscale objects such as nanoparticles,[38,61,63] NV centers,[39,40] erbium-doped nanocrystals,[44,45] proteins,[69,70] or virus.[71] Beyond the interest for quantum applications and plasmonic nanotweezers, our system can also be valuable for the spectroscopic studies of individual quantum dots,[51,52] releasing the need for surface tethering.

## ASSOCIATED CONTENT

**Supporting Information**

The Supporting Information is available free of charge on the ACS Publications website at DOI:xxx Additional method details, Nanogap antenna design, Numerical simulations of the intensity enhancement, Influence of the antenna parameters on the local intensity, Influence of the QD presence on the intensity enhancement, Infrared laser illumination influence on the photoluminescence emission, Two-photon absorption of the infrared laser, Temperature increase in the nanogap antenna, Comparison of trap stiffness in the gap antenna and double nanohole, Thermophilic response of the QDs in presence of SDS, Blinking analysis of single quantum dots, Photon antibunching of selected single quantum dot, Photoluminescence lifetime data.

**Funding Sources**

This project has received funding from the European Research Council (ERC) under the European Union's Horizon 2020 research and innovation programme (grant agreements No 723241 TryptoBoost) and from the Agence Nationale de la Recherche (ANR) under grant agreement ANR-17-CE09-0026-01.

**Conflict of Interest**

The authors declare no competing financial interest.

# Supporting Information for

# Single Photon Source from a Nanoantenna-Trapped Single Quantum Dot

Quanbo Jiang,[1] Prithu Roy,[1] Jean-Benoît Claude,[1] Jérôme Wenger[1,*]

[1] Aix Marseille Univ, CNRS, Centrale Marseille, Institut Fresnel, 13013 Marseille, France

* Corresponding author: jerome.wenger@fresnel.fr

This document contains the following supporting information:

S1. Additional method details

S2. Nanogap antenna design

S3. Numerical simulations comparing with other antenna designs

S4. Influence of the antenna parameters on the local intensity

S5. Influence of the QD presence on the intensity enhancement

S6. Infrared laser illumination influence on the photoluminescence emission

S7. Two-photon absorption of the infrared laser

S8. Temperature increase in the nanogap antenna

S9. Comparison of trap stiffness in the gap antenna and double nanohole

S10. Thermophilic response of the QDs in presence of SDS

S11. Blinking analysis of single quantum dots

S12. Photon antibunching of selected single quantum dots

S13. Photoluminescence lifetime data

## S1. Additional method details

*Nanoantenna fabrication*

First, a stack of 100 nm-thick gold with 5 nm-thick chromium as adhesion layer is deposited on top of glass coverslips using electron-beam evaporation (Bühler Syrus Pro 710). Nanoantennas are then milled using a gallium focused ion beam (FEI Strata DB235) with 30 kV acceleration voltage and 10 pA as ion beam current. The milling process is divided within two parts: first, carving the area surrounding the nanoantenna, the gap antenna is milled. Using this strategy, the gap size can be tuned easily.

*Quantum dots*

The quantum dots (QDs) used as objects for the trapping experiments are purchased from Thermo Fisher Scientific (Ref: Q21321MP). The QDs are carboxylate functionalized with 11 nm diameter. The QDs are made from nanometer-scale crystals of CdSe, which are shelled with an additional ZnS layer to improve the chemical and optical properties. The QDs have a narrow, symmetric emission band with the maximal peak at 655 nm excited at 635 nm. From the data provided by the supplier, the extinction coefficient at 635 nm is 800,000 cm$^{-1}$.M$^{-1}$. The quantum yield is determined by comparing the PL brightness with Alexa Fluor 647 reference, a 60% quantum yield is determined for this sample, in agreement with earlier reports.[1] The intensity-averaged PL lifetime is 35 ns.

The core-shell QDs are further coated with a polymer layer (containing $-COO^-$ surface groups) that allows the dispersion in aqueous solution with retention of their optical properties. The QDs are diluted to a final concentration of 2 nM with the ultrapure water with conductivity 18.2 MΩ.cm (Merck Millipore DirectQ-3 UV). 2 mM Sodium dodecyl sulfate (SDS) from Sigma Aldrich is used as a surfactant to avoid aggregation and sticking on the metal surface during the trapping experiment. No salts are added to the solution. To avoid the thermally-induced convection which may disturb the trapping experiment, the height of the solution of QDs on the top of the gap antenna is limited around 20 µm.[2]

For the reference sample where the QDs are fixed on the glass substrate, we passivate the clean borosilicate glass surface by a positively-charged poly-L-lysine layer so as to electrostatically bind the negatively-charged QDs. The 0.01% poly-L-lysine solution (Sigma Aldrich P4707) is dropped on the glass and incubated for 5 minutes at room temperature before rinsing with water. The sample is then dried at 37°C for 2 hours. Lastly, the diluted solution containing the QDs at 200 pM is deposited on the polylysine-coated surface and further rinsed with water. Without this surface treatment, no QD remains on the glass surface.

*Experimental setup*

The inverted confocal microscope is developed for plasmonic nano-optical trapping with a continuous wave 1064 nm laser (Ventus 1064-2W). A 635 nm pulsed laser diode (Picoquant LDH-P-635) at 5 MHz repetition rate is aligned with the infrared laser to excite the photoluminescence of quantum dots. Both lasers are focused by a high NA microscope objective (Zeiss Plan-Neofluar 40x, NA 1.3, oil immersion) and their spot diameters are measured as 1 µm and 0.6 µm for the 1064 and 635 nm lasers, respectively. The infrared trapping intensity at the objective focus is directly expressed in mW/µm² considering the 50% transmission of the objective at 1064 nm. The photoluminescence emission from QDs is collected by the same microscope objective in epi- configuration. A set of dichroic mirrors, long pass filters, 30 µm confocal pinhole and bandpass filters ensures that only the photoluminescence signal is detected without the noise from the scattered light by any of the two lasers. Two avalanche photodiodes APDs (Picoquant MPD-5CTC) separated by a 50/50 beam-splitter in a Hanbury-Brown-Twiss configuration record the emitted photons in the spectral range of 650-700 nm. The photodiode outputs are connected to a time correlated single photon counting (TCSPC) module (Picoquant Picoharp 300).

The overall detection efficiency of the setup is estimated at 3% as the product of the 30% microscope objective collection efficiency, the 80% transmission of the microscope objective, the 70% transmission of our optics (including the reflectivity of the silver mirrors), the 65% transmission of the filter stack and the 30% quantum efficiency of our APDs.

*Trap stiffness quantification*

Following the approach detailed in [3–7], the temporal correlation of the PL intensity recorded during a trapping event decay as $\exp(-2t/\tau)$, where the characteristic time $\tau = \gamma/\kappa$ is the ratio of the Stokes drag coefficient $\gamma$ by the trap stiffness $\kappa$. The Stokes drag coefficient $\gamma$ which is given by Faxen's law to account for the influence of the interface near the nanoparticle:[8,9]

$$\gamma = \frac{6\pi\eta R}{\left(1 - \frac{9}{16}\frac{R}{h} + \frac{1}{8}\left(\frac{R}{h}\right)^3 - \frac{45}{256}\left(\frac{R}{h}\right)^4 - \frac{1}{16}\left(\frac{R}{h}\right)^5\right)} \quad (S1)$$

where $R$ is the nanoparticle radius, $\eta$ the viscosity of the water medium and $h$ is the average distance between the center of the nanoparticle and the dimer aperture border. In our trapping experiment, we set $R = 5.5$ nm and $h = 11$ nm for the gap size of 22 nm and $h = 17$ nm for the gap size of 34 nm. We also take into account the temperature dependence of the water viscosity $\eta$ using the Vogel equation [9] and our calibration of the temperature inside the gap antenna as a function of the infrared laser power.

*Numerical simulations*

Frequency domain study from Optics module of COMSOL Multiphysics 5.5 is used for simulation of the nanostructures. The software is installed on Dell Precision Tower 5820 with 64GB RAM and 8-core Intel Xeon processor for fast convergence of solutions. The user-defined tetrahedral mesh is used with mesh size ranging from 0.1 nm (in the gap and region of interest) to 10 nm (at the boundaries of model in the surrounding media). The plane wave excitation with electric field value of 1V/m is used. The simulation domain is bounded by PML (perfectly matched layer) to suppress any reflection from the domain boundaries. Intensity distribution maps are represented in a plane 5 nm above the glass-gold interface. The vertical profile of the simulated nanoantenna takes into account the tapering due to FIB milling. While the gap is 22 nm at the glass-gold interface, it tapers up to 122 nm at the top of the antenna (at the maximum height of the 100 nm-thick gold film). To mimic our experiments, light is incoming from the bottom of the nanoantenna where the gap is the shortest. To account for the presence of the QD in our simulations, we add a 10 nm diameter sphere of 2.47 refractive index in the center of the nanogap.

**S2. Nanogap antenna design**

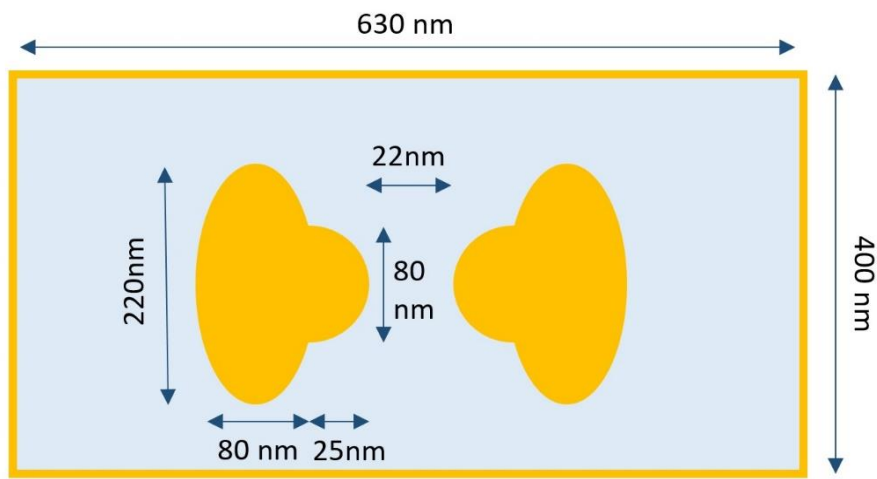

**Fig. S1**. Geometrical parameters of our nanogap antenna milled in gold.

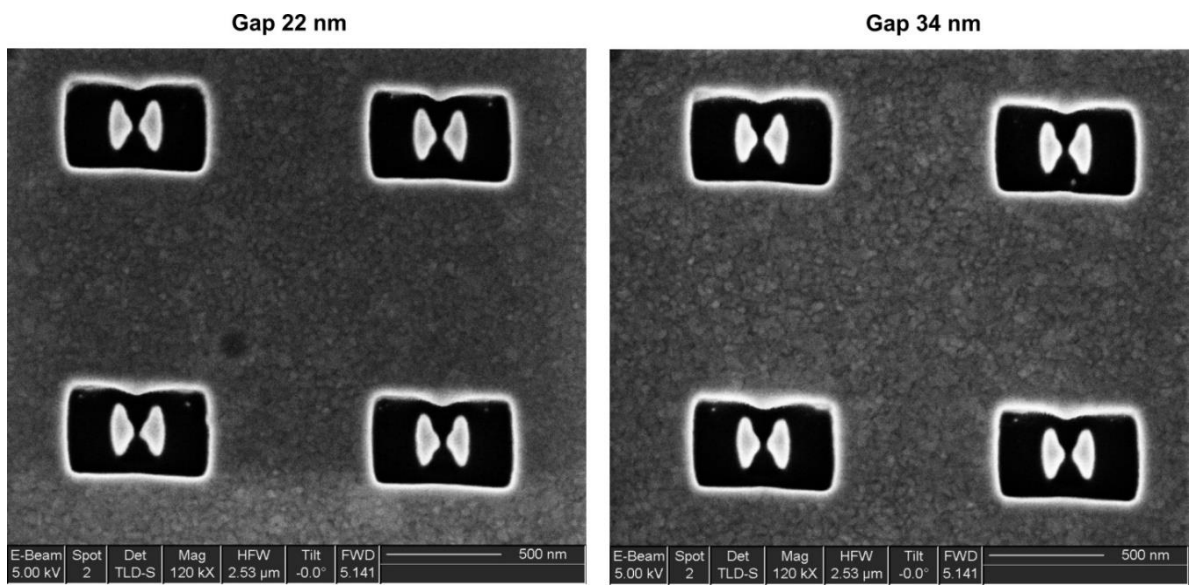

**Fig. S2**. Scanning electron microscopy images of nanogap antennas fabricated by focused ion beam with 22 or 34 nm gap.

## S3. Numerical simulations comparing with other antenna designs

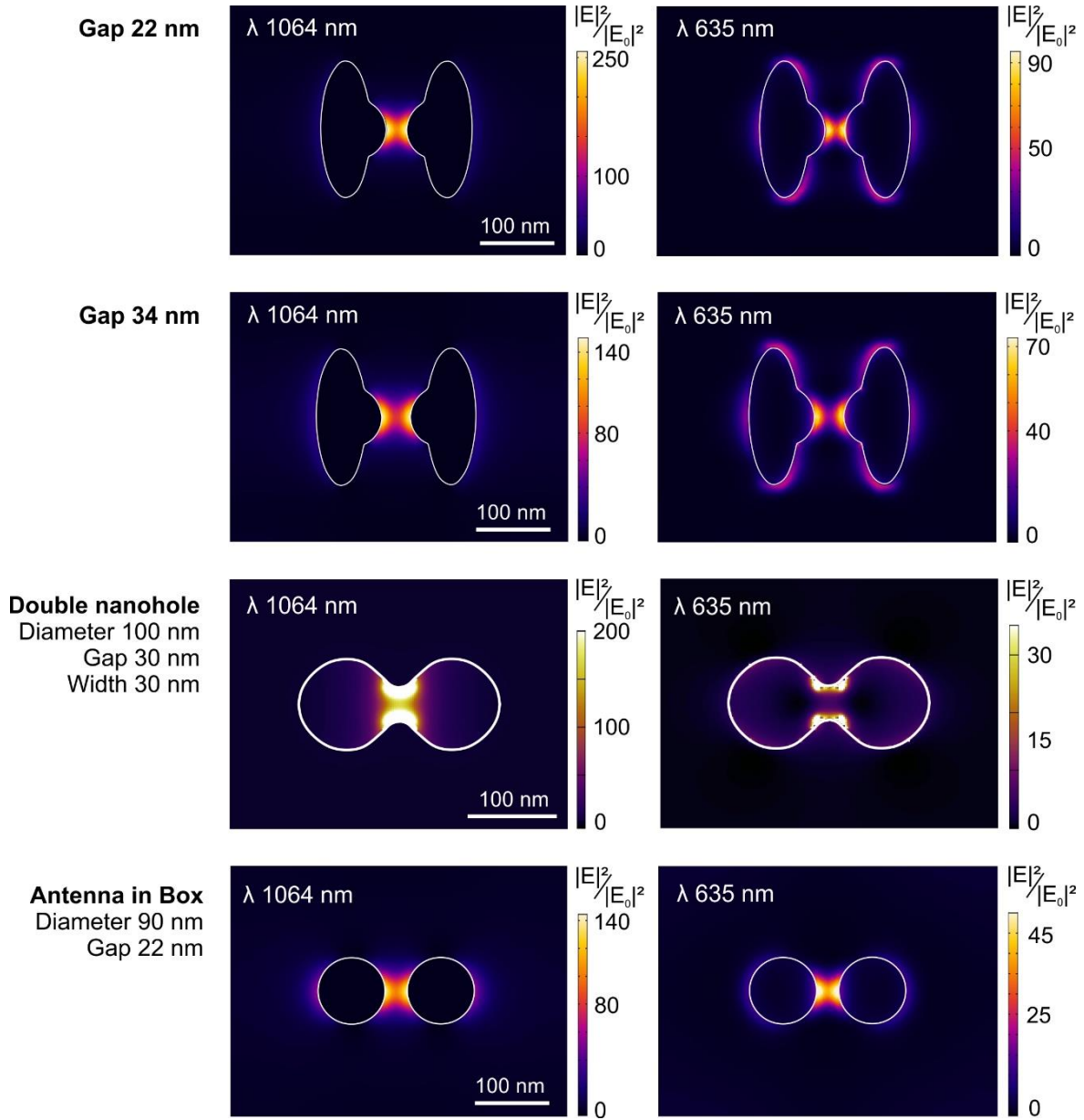

**Fig. S3**. Comparison of the intensity enhancement in different nanoantenna designs at 1064 nm and 635 nm illumination. The intensity distribution is shown in a plane 5 nm above the gold-glass interface and the presence of the QD is not taken into account here. An improved performance is obtained with the 22 nm gap as compared to the 34 nm gap antenna. While the optimized double nanohole design yields a slightly higher intensity enhancement at 1064 nm,[7] its lower performance at 635 nm does not enable observing a clear PL enhancement as compared to the glass reference. The circular pattern of the antenna in box is conceptually simpler, yet we found that the additional hotspots present at the extreme edges of the dimer were leading to multiple QD trapping events. The elongated shape of our antenna design allows to reduce the intensity at the extreme edges without affecting the central hotspot in the gap region.

## S4. Influence of the antenna parameters on the local intensity

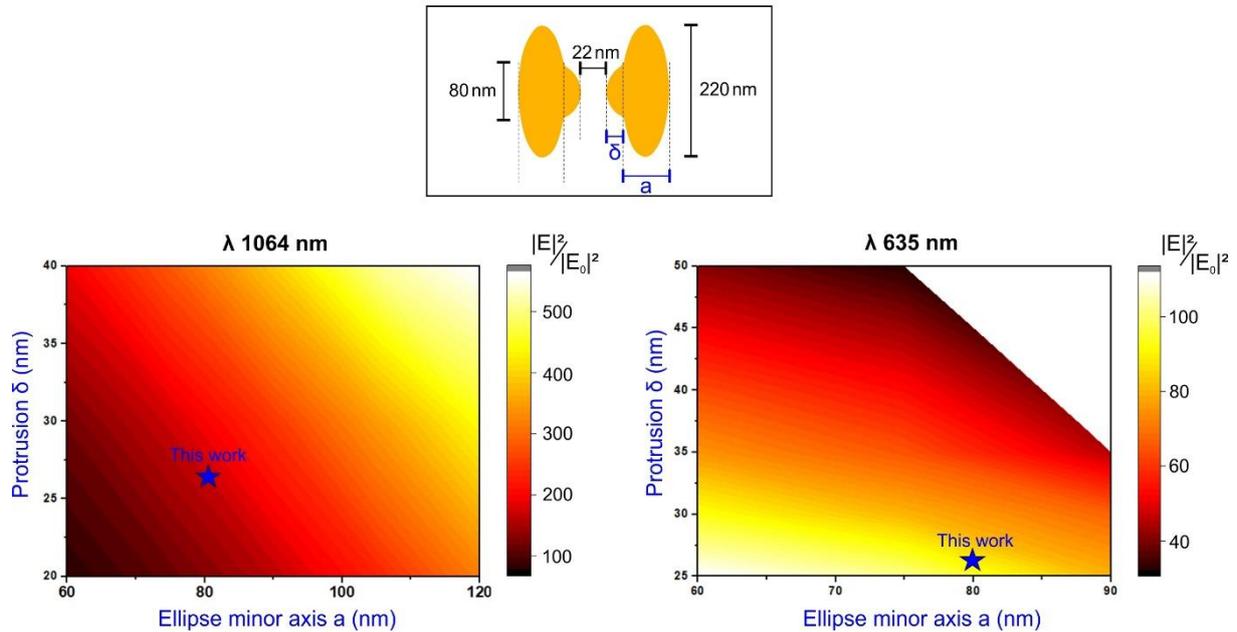

**Fig. S4**. Influence of the antenna parameters on the intensity enhancement in the center of the gap at 1064 and 635 nm. The sketch on the top indicates the dimensions used here (black numbers) and the two parameters explored (blue labels), namely the short axis of the ellipse a and the size of the protrusion δ. The presence of the QD is not taken into account here. Due to the large spectral shift between the two excitation wavelengths, there is no single choice (a, δ) maximizing simultaneously the antenna enhancements at the two wavelengths. A compromise has to be found depending whether more emphasis is desired to maximize the 635 nm enhancement (this work) or the 1064 nm enhancement (to further improve the trap stiffness). Two main reasons contribute to explain the higher enhancement at 1064 nm as compared to 633 nm: (i) the plasmonic losses are about twice lower in the infrared and (ii) the apparent gap size is λ/50 for the infrared while it is λ/30 for the red.

## S5. Influence of the QD presence on the intensity enhancement

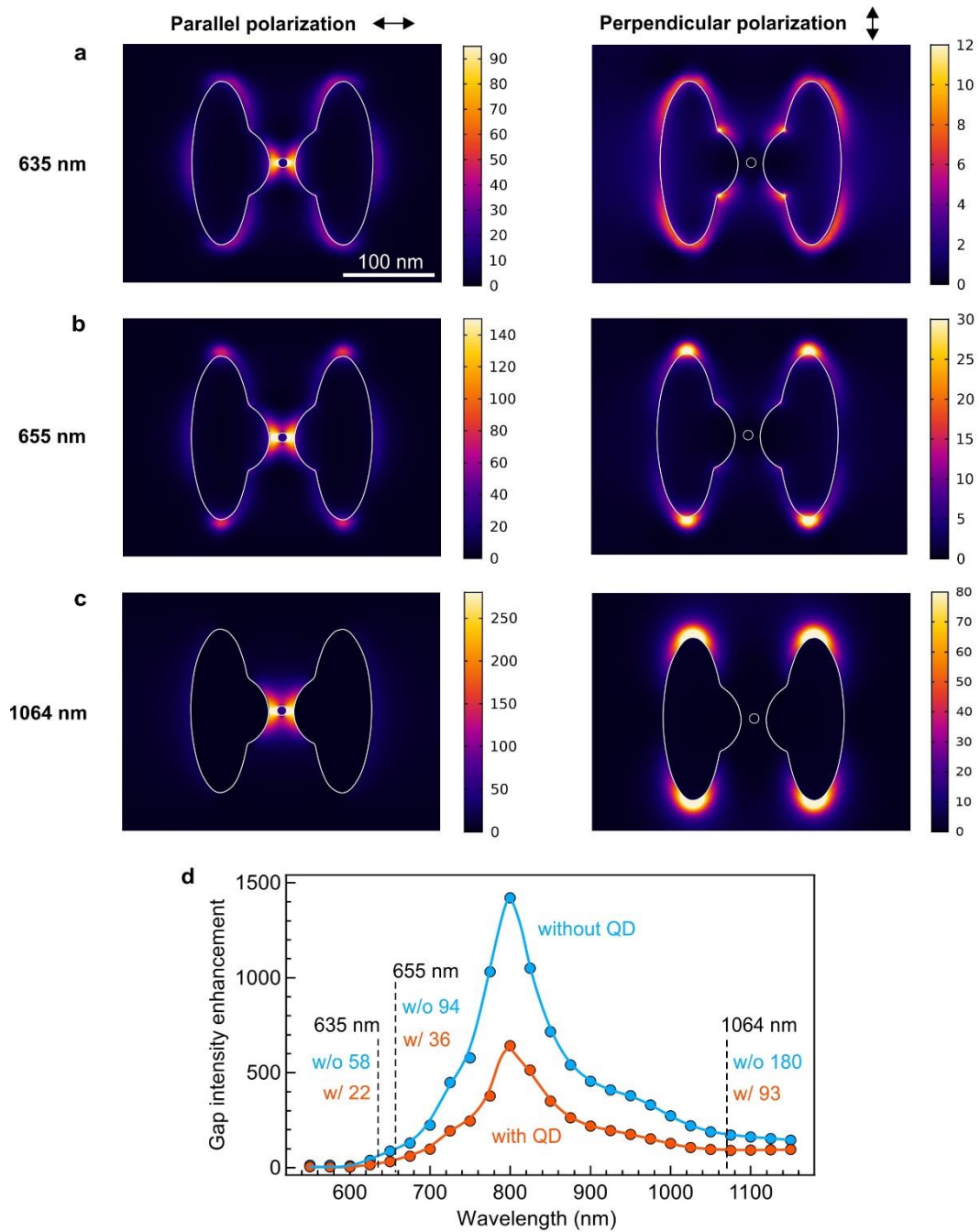

**Fig. S5**. Influence of the presence of the QD on the local intensity enhancement $|E|^2/|E_0|^2$. The QD is modeled as a 10 nm sphere of 2.47 refractive index and the gap is 22 nm. (a-c) Intensity distributions for different wavelengths and incident polarizations. (d) Intensity enhancement in the center of the nanogap as a function of the incident wavelength. When the QD is taken into account (orange curve), the data is normalized by the intensity in the center of the 2.47 refractive index sphere in the absence of the nanoantenna (~0.7× the incident intensity due to the scattering by the nanosphere). For the other case (blue curve), the data is normalized by the incoming intensity. Due to the presence of the QD, the spectral resonance of the nanoantenna is redshifted from 800 to 806 nm. Depending on the wavelength, the presence of the QD reduces the intensity by 2 to 3× as compared to the reference simulations without the QD.

## S6. The infrared laser illumination does not affect the photoluminescence emission and diffusion of quantum dots

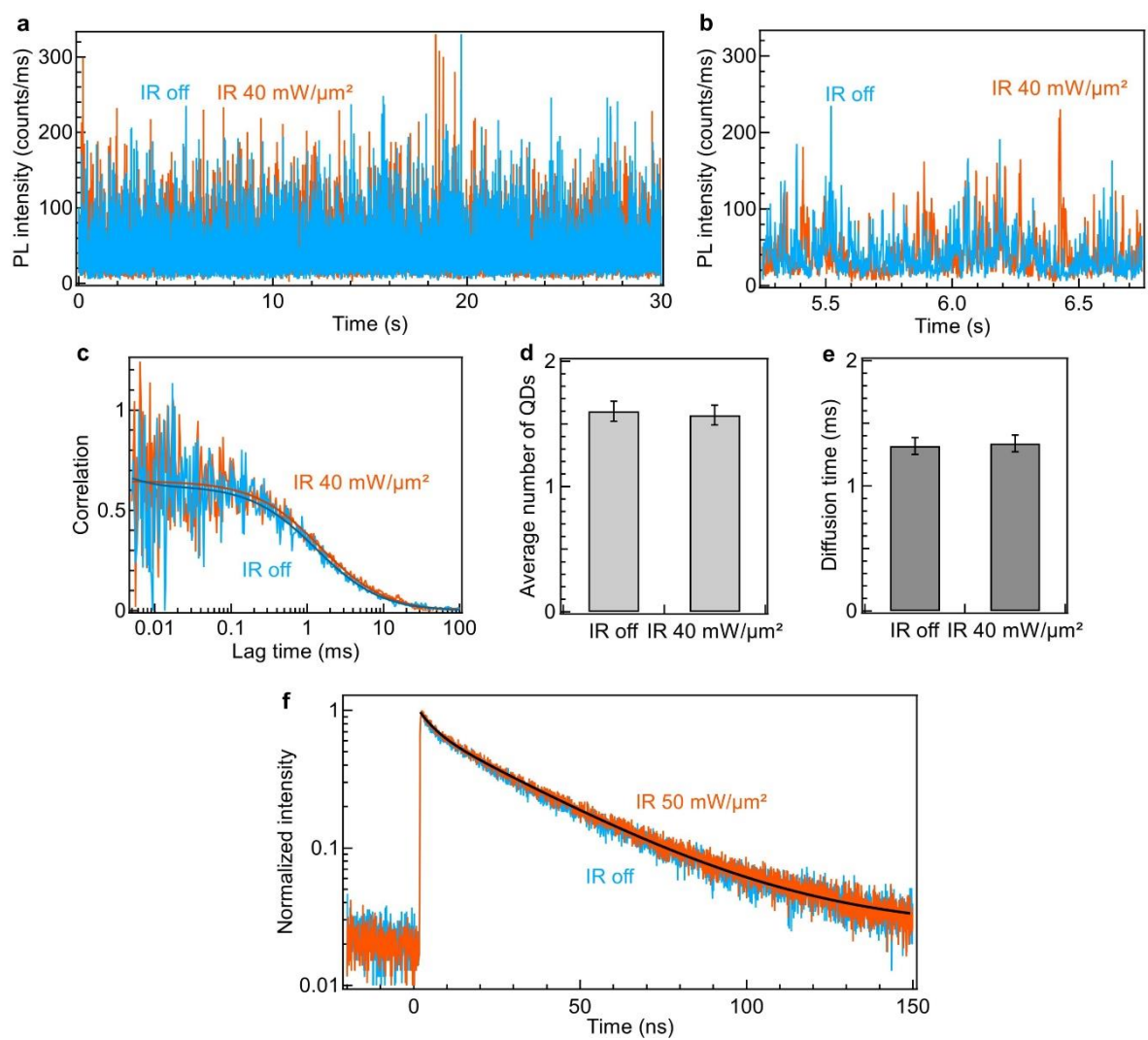

**Fig. S6**. (a) Photoluminescence intensity time traces of free-diffusion quantum dots measured in the confocal volume with and without the infrared laser. The red excitation laser power is kept constant at 10 µW. (b) Zoom-in of the time traces of (a), showing no clear change of the QD signal in presence of 40 mW/µm² illumination at 1064 nm. (c) Fluorescence correlation spectroscopy analysis of the time traces in (a). From the numerical interpolation of this data based on a standard 3D Brownian diffusion model,[10] the average number of quantum dots in the confocal detection volume (d) and their diffusion time (e) are deduced. Both the number of quantum dots and their diffusion time indicate no influence of the presence of the high-power infrared laser.[6] (f) PL lifetime histograms recorded with and without the 1064 nm illumination. No significant change is seen on the PL lifetime decay.

## S7. The two-photon absorption of the infrared laser by the quantum dots remains negligible

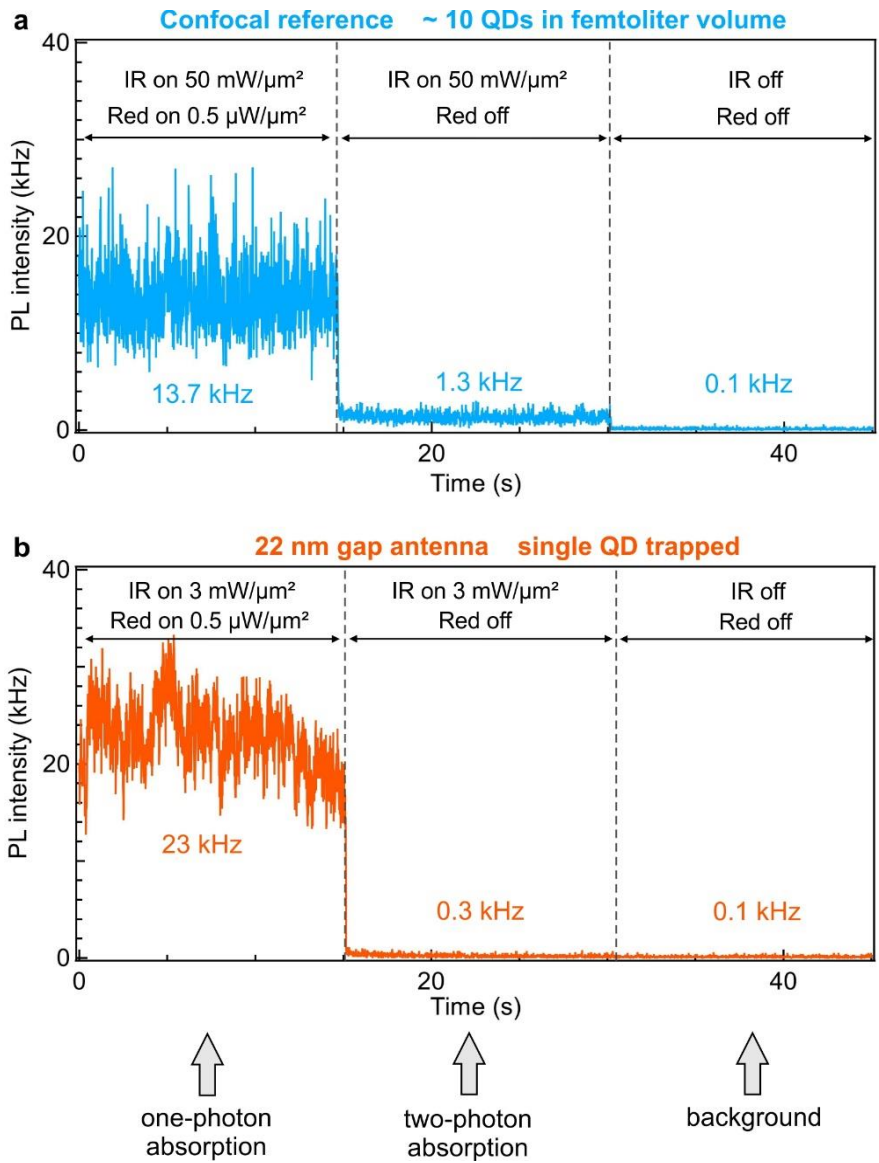

**Fig. S7**. Measurement of the two-photon absorption (TPA) of the infrared trapping laser exciting the QD luminescence. For the first 15 s of the experiment, both lasers are turned on and the PL signal is dominated by the one-photon excitation of the QDs. Then the red laser is blocked and only the IR laser illuminates the sample. Finally at t > 30 s, both lasers are blocked to record the background level. Experiments in (a) are performed on a 20 nM solution of QDs so that approximately 10 QDs are detected in the diffraction-limited confocal volume. In (b), we use a 22 nm gap antenna and ensure a single QD is trapped. The average PL intensity levels are indicated in each time interval. For the confocal reference, we estimate a TPA signal of 0.12 kcts/s per QD at 50 mW/µm² infrared intensity, which would correspond to 0.4 counts per second at 3 mW/µm². Once the background is subtracted, our observation of 200 counts per second for the TPA of the antenna-trapped QD indicates a 500× enhancement of the TPA emission in the antenna. Although this large value highlights the performance of our antenna, the total TPA signal in the antenna does not exceed 1% of the one-photon PL excited by the 635 nm laser.

## S8. Temperature increase in the nanogap antenna

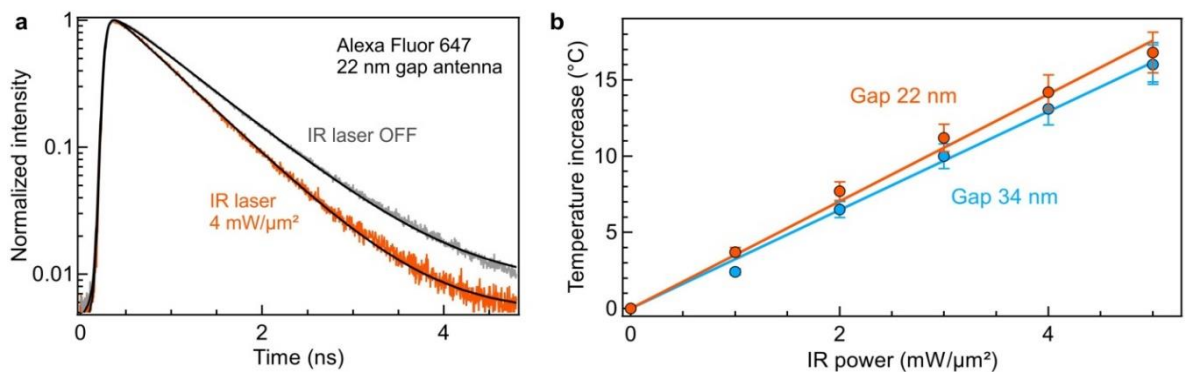

**Fig. S8**. (a) Normalized fluorescence decay and fitting curves of Alexa Fluor 647 recorded in the 22 nm gap antenna with and without infrared illumination. The shortening of the Alexa dye lifetime is due to the temperature increase consecutive to the infrared absorption in the plasmonic nanoantenna.[11,12] (b) Temperature increase as a function of the infrared power based on the lifetime measurement in the nanoantennas with 22 nm and 34 nm gap sizes, following the method described in [11,12]. As already observed for double nanoholes,[11] the temperature increase is set by the absorption in the metal film, and is largely independent of the gap size, plasmon resonance or illumination polarization.

## S9. Comparison of trap stiffness in the gap antenna and double nanohole

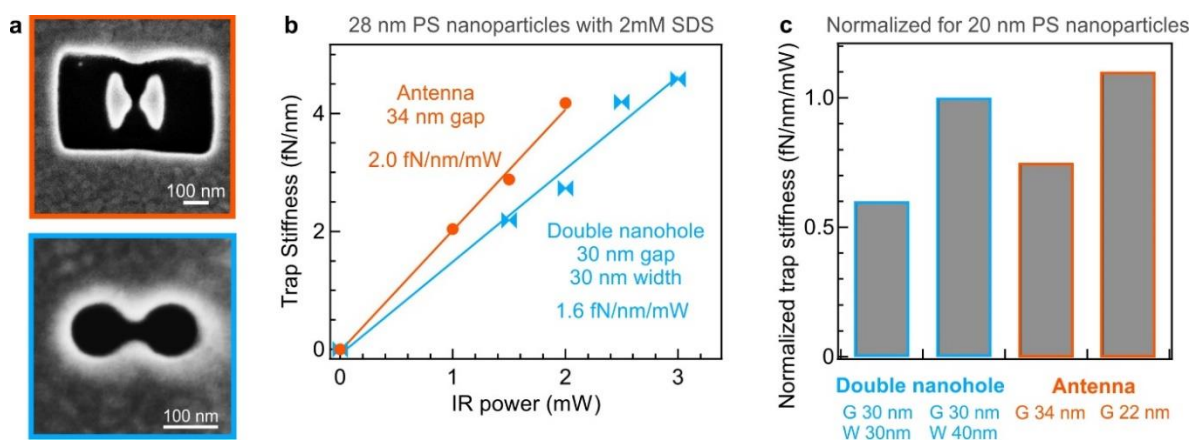

**Fig. S9**. (a) Scanning electron microscopy image of the gap antenna with 34 nm gap and the double nanohole with 30 nm gap. (b) Trap stiffness of 28 nm-diameter polystyrene nanoparticles as a function of the infrared power. The data for the double nanohole is taken from ref.[6], and 2 mM SDS is used as surfactant in both cases. The higher trap stiffness with the nanogap antenna demonstrates its superior performance. (c) Normalized trap stiffness computed for 20 nm-diameter polystyrene nanoparticles derived from our experimental data. This data accounts for the polarizability dependence with the nanoparticle diameter and refractive index.

## S10. Thermophilic response of the QDs in presence of SDS

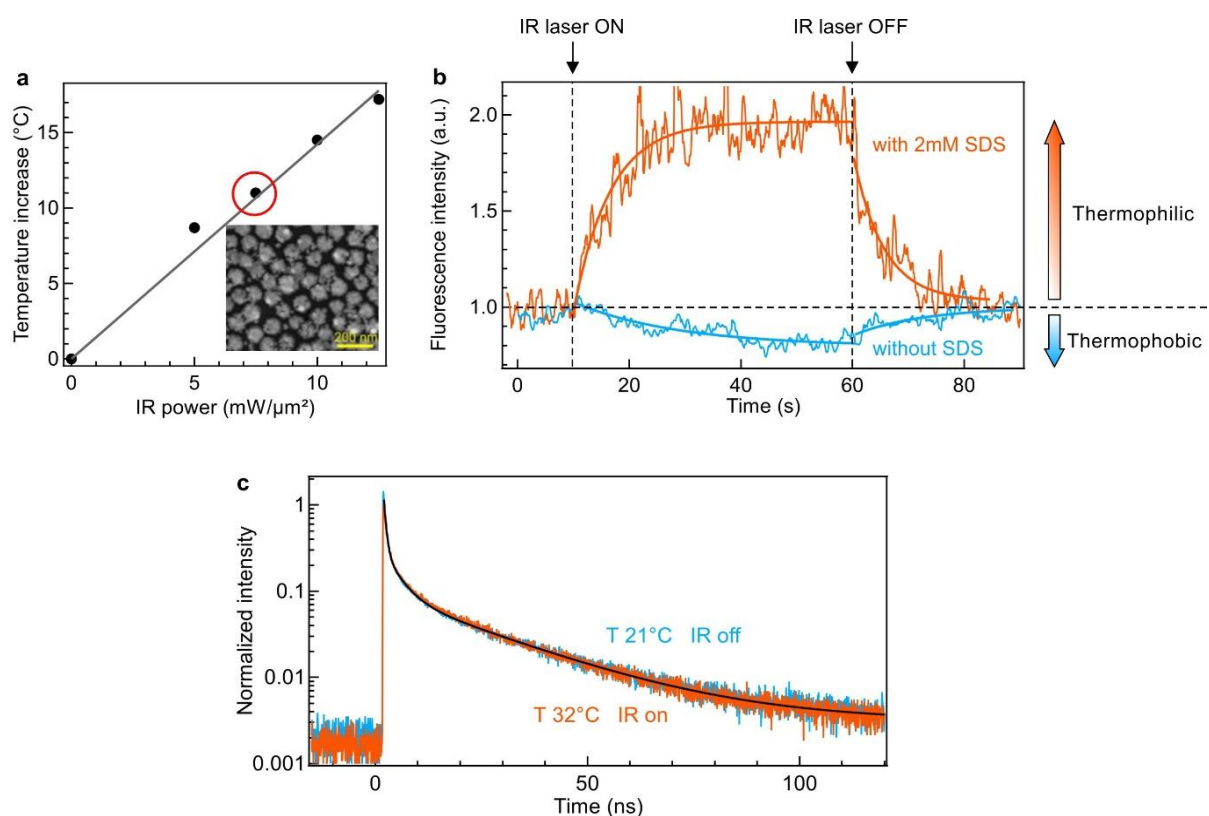

**Fig. S10**. For this set of experiments, we use a glass coverslip covered with nanoparticles to allow recording a larger signal from an ensemble of QDs in solution. (a) Calibration of the temperature increase as a function of the infrared power. The point in the red circle reproduces the temperature increase in the gap antennas used for the trapping experiment (Fig. S7). (b) Photoluminescence intensity evolutions of quantum dots with and without 2 mM SDS surfactant. The infrared laser is switched on from 6 s to 60 with 7 mW/μm² corresponding to the temperature increase indicated in the red circle in (a). The increase and drop of photoluminescence intensities represent the thermophilic and thermophobic behaviors of quantum dots, respectively in presence of a temperature gradient. In the thermophilic condition (with SDS), the thermophoretic force contributes to promote the plasmonic optical trapping.[6] (c) PL lifetime histograms extracted from the time trace in (b) using 2 mM SDS when the temperature is stable at +32°C (time interval between 30 and 60 s in (b) with the IR laser turned on) and when the sample goes back to room temperature (corresponding to 90 to 150 s in (b)). For the temperature range probed here, we do not monitor any noticeable change on the average PL lifetime depending on the local temperature. While the presence of the gold nanoparticles shortens the PL lifetime by ~2x, the relative test with and without infrared heating does not indicate any major dependence of the PL lifetime with the temperature.

## S11. Blinking analysis of single quantum dots

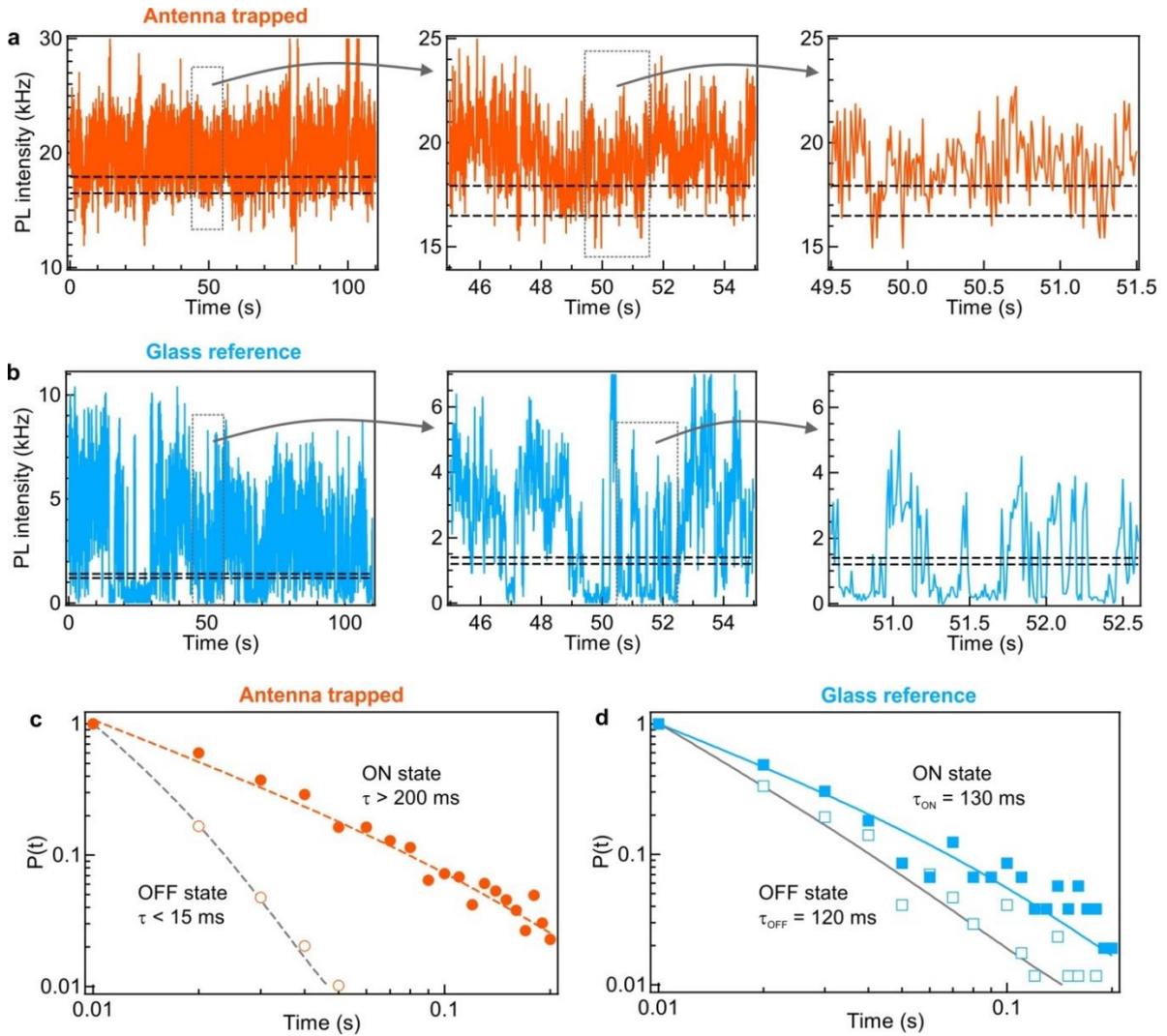

**Fig. S11**. (a,b) PL intensity time traces of a trapped quantum dot in the gap antenna and a fixed quantum dot on a glass coverslip. The on- and off- states are defined as the signal level higher than the top dashed line and lower than the bottom dashed line, respectively. While for the fixed QD the levels are easy to define, the strongly suppressed blinking in presence of the nanogap antenna complicates the discrimination between states. Our definition of levels take into account the statistical noise as shown by the insets in (a). (c,d) Single QD probabilities as a function of the dwell time. The on- and off-state probabilities $P(t)$ are defined as the number of on or off events of duration time divided by the total number of on or off events.[13] The probabilities are fitted with a truncated power law[14–16] as $P(t) = Ax^{-m}e^{-(x/tau)} + y_0$ where $A$ is the amplitude, m is the power law exponent and $tau$ is the characteristic blinking time which is noted in (c,d). While on the glass coverslip the QD spends similar times in the on- and off- state, the nanoantenna significantly promotes the duration of the on- state events and reduces the off-state probability, quenching the blinking dynamics.

**S12. Photon antibunching of selected single quantum dots**

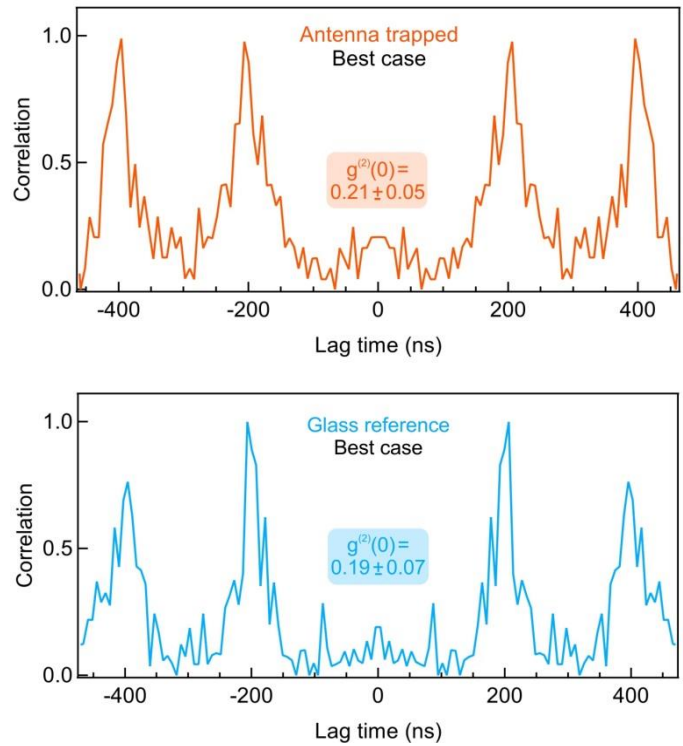

**Fig. S12**. Photon antibunching data of selected single quantum dots trapped in the gap antenna and fixed on a glass coverslip. Strong antibunching with $g^{(2)}(0) \approx 0.2$ are observed in both cases, which constitute our best antibunching result. However, this strongly depends in the chosen QD, and was observed for less than 10% of all the QDs investigated. No clear correlation between the QD brightness and the $g^{(2)}(0)$ value could be evidenced. Nevertheless, this data demonstrates that the residual correlation found at zero lag times depends on the properties of the QD source (we use here commercially-available QDs), and are not limited by the nanoantenna trap.

## S13. Photoluminescence lifetime data

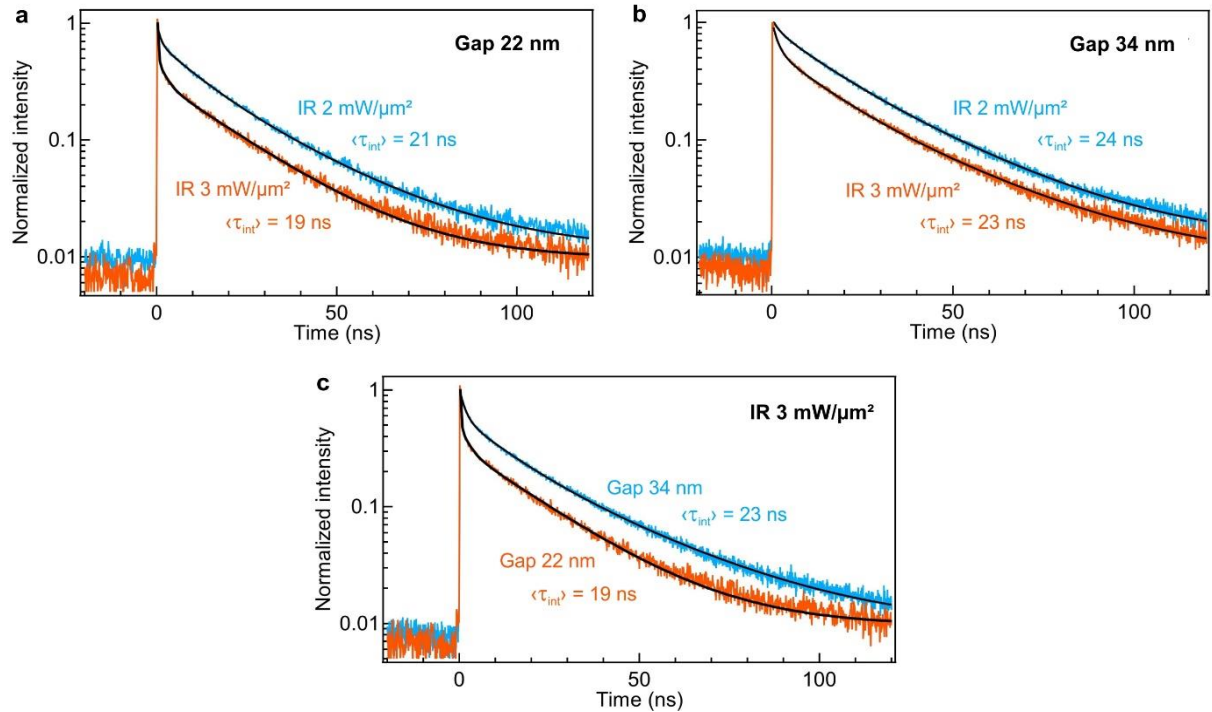

**Fig. S13**. PL lifetime histograms for different trapping experiments with increasing IR powers on antennas with 22 nm gap (a) and 34 nm gap (b). (c) Comparison between the PL decays for 22 and 34 nm gap antennas at a fixed 3 mW/µm² IR intensity. The lifetimes indicated on the graphs are intensity-averaged. The numerical fit results of the triple-exponential model are detailed in Tab. S1 below.

**Table S1.** Fit parameters for the PL lifetime data displayed on Fig. 3h and S13. The lifetimes are expressed in ns, $A_i$ denote the amplitudes. The FWHM of our instrument is 110 ps. We use a biexponential fit for the QD on glass reference, while we find that a triple-exponential fit best describes the PL decay for the QD trapped in the nanoantenna.

|  | $\tau_1$ | $\tau_2$ | $\tau_3$ | $A_1$ | $A_2$ | $A_3$ | Intensity-averaged lifetime |
|---|---|---|---|---|---|---|---|
| Glass reference | - | 7.7 | 38 | - | 0.28 | 0.65 | 35 |
| 22 nm gap antenna IR 2 mW/µm² | 0.9 | 10 | 25 | 0.35 | 0.49 | 0.65 | 21.1 |
| 22 nm gap antenna IR 3 mW/µm² | 0.3 | 2.7 | 20.4 | 0.58 | 0.36 | 0.8 | 19.2 |
| 34 nm gap antenna IR 2 mW/µm² | 2 | 14 | 29 | 0.9 | 2.5 | 2.9 | 24.2 |
| 34 nm gap antenna IR 3 mW/µm² | 1.5 | 12 | 28 | 1.9 | 1.4 | 1.8 | 23.0 |